# The quantum mechanics *needs* the principle of wave-function collapse – but this principle shouldn't be *misunderstood*

*by Sofia D. Wechsler*

**Abstract**
The postulate of the collapse of the wave-function stands between the microscopic, quantum world, and the macroscopic world. Because of this intermediate position, the collapse process cannot be examined with the formalism of the quantum mechanics (QM), neither with that of classical mechanics. This fact makes some physicists to propose interpretations of QM, which avoid this postulate. However, the common procedure used in that, is making assumptions incompatible with the QM formalism. The present work discusses the most popular interpretations. It is shown that because of such assumptions those interpretations fail, i.e. predict for some experiments results which differ from the QM predictions.
Despite of that, special attention is called to a proposal of S. Gao, the only one which addresses and tries to solve an obvious and major contradiction.
A couple of theorems are proved for showing that the collapse postulate is *necessary* in the QM. Although non-explainable with the quantum formalism this postulate cannot be denied, otherwise one comes to conclusions which disagree with the QM.
It is also proved here that the idea of 'collapse at a distance' is problematic especially in relativistic cases, and is a misunderstanding. Namely, in an entanglement of two quantum systems, assuming that the measurement of one of the systems (accompanied by collapse of that system on one of its states) collapses the other system too without the second system being measured, leads to a contradiction.

**Abbreviations**
CSL     = continuous spontaneous localization
dBB     = de Broglie-Bohm
EPR     = Einstein, Podolsky, Rosen
F&H     = Feynman and Hibbs
GPR     = Ghirardi, Pearle, Rimini
GRW     = Ghirardi, Rimini, Weber
LHS     = left hand side
QM      = quantum mechanics
RHS     = right hand side
SSE     = stochastic Schrödinger equation
w-f     = wave-function
w-p     = wave-packet

## 1. Introduction

In a profound analysis of tests of quantum systems, J. von Neumann concluded that once a quantum system in the initial state $\psi$ is tested (non-destructively) and produces the result $\lambda_k$, the system remains in a state $\phi_k$ with the property that any subsequent measurement of the system for the same observable, would produce the same result, $\lambda_k$ (see for instance page 138 in [1] ). This conclusion was always confirmed by the experiment.

G. Lüders refined von Neumann's work examining cases with degenerate eigenvalues [2]. But the question remained what happens with the other components $\phi_1, \ldots, \phi_{k-1}, \phi_{k+1}, \ldots, \phi_N$ of the initial wave-function (w-f). Do they disappear, or do they continue to exist?

Complementarily, if a test is done for selecting the result $\lambda_k$ and the detector remains silent, was the wave-packet (w-p) $\phi_k$ destroyed?

Some scientists believe that $\phi_1, \ldots, \phi_{k-1}, \phi_{k+1}, \ldots, \phi_N$ in the first case, and $\phi_k$ in the second case, do disappear from the world. But neither von Neumann, nor Lüders, brought a rigorous proof to that. So, the idea of reduction of the w-f, or 'collapse', remained just as a postulate. Lüders motivated:

> "statements on the change of state due to measurement do not arise out of quantum theory itself through the inclusion of the measurement apparatus in the Schrödinger equation. Measurement, an act of cognizance, adds an element not already contained in the formulation of quantum theory."

It is not clear what Lüders meant by the words 'act of cognizance', because a measurement can be done with no human presence. About the measurement apparatus, indeed, it cannot be included in the quantum treatment since it is a macroscopic apparatus, not a quantum one.

Other physicists, displeased by the enigmatic collapse postulate, launched 'interpretations' of the quantum mechanics (QM). In general, for explaining the measurement process of quantum systems without this postulate, these interpretations introduce modifications in the standard quantum formalism, or even in physical laws. Sometimes the modifications seem very logical, though, there is always a price to be paid, a contradiction with some quantum predictions, or, self-contradictions. So happens with the most popular interpretations, e.g. the mechanics of de Broglie and Bohm [3, 4], the full/empty waves hypothesis (see for example [5] for explanation of the concept), Gao's jumping particles [6], the consistent histories [7], the transactional interpretation [8], Everett's many-worlds [9]. The section 2 will exemplify in more detail advantages and weaknesses of these interpretations.

13 years after the two articles of Bohm presenting his interpretation of QM, Feynman and Hibbs (F&H) explained in their book "*Quantum Mechanics and Path Integral*", [10], the reduction of a w-p as an effect of increasing of the system in dimensions (size, mass, etc.). In the subsection 2.3 of the book they wrote,

> "The classical approximation, however, corresponds to the case that the dimensions, masses, times, etc., are so large that S is enormous in relation to $\hbar$ (=1.05 × 10⁻²⁷ erg·sec). Then the phase of the contribution S/$\hbar$ is some very, very large angle ... small changes of path will, generally, make enormous changes in phase, ... The total contribution will then add to zero; for if one path makes a positive contribution, another infinitesimally close (on a classical scale) makes an equal negative contribution. ... But for the special path $\bar{x}(t)$, for which S is an extremum, a small change in the path produces, in the first order at least, no change in S. All the contributions from the paths in this region are nearly in phase, ..., and do not cancel out"[1]

Except the de Broglie and Bohm (dBB) interpretation of the QM, the other interpretations were proposed after the book of F&H. Though, all those interpretations ignored the physical and rigorous explanation in [10]. The same is true about the dBB interpretation. One reason may justify this attitude: F&H restricted their explanation to a w-f consisting in one single w-p. To generalize that explanation to a superposition of a couple of w-ps is not simple, because the F&H rationale was done on all the possible paths between two given space-time points.

---

[1] S is the action function.

W-ps traveling in separated regions of the space have no common final space-time point. Even more complicated is the situation when the w-f is an entanglement.

There are though two interpretations which, to the difference from the above ones, took the collapse 'seriously' and suggested formalisms meant to bridge between the quantum and the classical world. Ghirardi, Rimini and Weber (GRW) [11] thought that the w-f of a quantum system might undergo at random times a sudden *shrinking* to a small region (localization). A few years later, Ghirardi, Pearle, and Rimini (GPR) came with a modified version of the GRW interpretation, the 'continuous spontaneous localization' (CSL) [12], by which the localization occurs progressively in time, instead of suddenly (see also [13] sections 7 and 8).
These interpretations also introduced changes in the Schrödinger equation, but the effect of the changes is negligible if the quantum system contains a small number of components, and is great – localization – when very many components are involved. Thus, the quantum formalism would not be changed for systems belonging to the quantum world, so that, in fact, these are not interpretations of the QM but models of collapse.

A noticeable remark capturing the situations in which the collapse occurs, is due to D. Bedingham [14]:

"Our experience in the use of quantum theory tells us that the state reduction postulate should not be applied to a microscopic system consisting of a few elementary particles until it interacts with a macroscopic object such as a measuring device."

The CSL model of collapse was successfully applied in [15] to the process in a detector, showing how the localization appears when more and more particles from the detector are involved.

*Note 1:* This text uses frequently the word 'particle', and it also appears in citations. Unless otherwise specified, this word means a single component quantum system, an electron, a proton, an atom or even a molecule whose internal structure is disregarded.

Since the collapse process cannot be explained using strictly the QM formalism, neither von Neumann, nor Lüders, considered to prove that this postulate is unavoidable. The present text contains proofs by which denying the collapse, one comes to contradictions with the QM predictions.

The rest of the text is organized as follows: section 2 discusses the most popular interpretations of QM which deny the collapse. Section 3 presents in brief the CSL model of collapse. Section 4 proves a couple of theorems in support of the necessity of the reduction postulate. Section 5 illustrates the fact that 'collapse at a distance', i.e. by virtue of the test of another system is a misunderstanding of the concept of collapse and leads to contradictions. Section 6 contains conclusions.

## 2. Criticism of some widespread interpretations of quantum mechanics

All the interpretations of the QM discussed below, ignore the well known experimental fact that the reduction of the w-f occurs, as F&H, Bedingham, GRW, and GPR said, in the presence of a macroscopic object that provides the very big number of particles necessary for the process.

### 2.1. The de Broglie-Bohm interpretation

This interpretation is based on the assumption that there exist substructure elements of the quantum world. One of them, named below 'dBB particle', floats inside the w-f, and moves as the w-f allows. Another basic assumption is that a dBB particle doesn't appear out of nowhere and doesn't disappear into nothing; therefore, it has a continuous trajectory which doesn't begin suddenly and doesn't stop suddenly, doesn't split into a couple of trajectories, and a couple of trajectories don't merge into one. A third assumption is that the trajectory admits the first time-derivative, i.e. at any point exists a velocity. That contradicts the uncertainty principle, which forbids the simultaneous existence of a well defined position and velocity for a quantum system. dBB tried to justify this contradiction by claiming that the measurement of position disturbs the velocity.

This interpretation shows a major weakness when applied to entanglements and relativistic situations, as found by Berndl et al. [16] in an analysis of the consequences of the famous Hardy's paradox [17]. A simpler explanation of the contradiction can be found in [18]. It is repeated below in general lines.

A pair of entangled particles, $p^+$ and $p^-$ are produced in a state entangled by paths – figure 1,

$$|\psi\rangle_0 = \frac{1}{\sqrt{3}} \left( i|u^+\rangle|v^-\rangle + |v^+\rangle|v^-\rangle + i|v^+\rangle|u^-\rangle \right). \tag{1}$$

The reader can check that after passing through the beam-splitters $BS^+$, respectively $BS^-$ the state of the two particles becomes

$$|\psi\rangle_1 = \frac{1}{2\sqrt{3}} \left( -3|c^+\rangle|c^-\rangle + i|c^+\rangle|d^-\rangle + i|d^+\rangle|c^-\rangle - |d^+\rangle|d^-\rangle \right). \tag{2}$$

The result which raises problems is $|d^+\rangle|d^-\rangle$ i.e. a joint click in the detectors $D^+$ and $D^-$.

Consider that the two particles are tested in different labs.

In a frame of coordinates $\mathcal{F}^+$ flying in the direction from the lab where is tested $p^-$ to the lab where is tested $p^+$, the time axis would show that when $p^+$ is detected, $p^-$ didn't yet meet the beam-splitter $BS^-$. The calculi in [18] show that the corresponding state of the system is

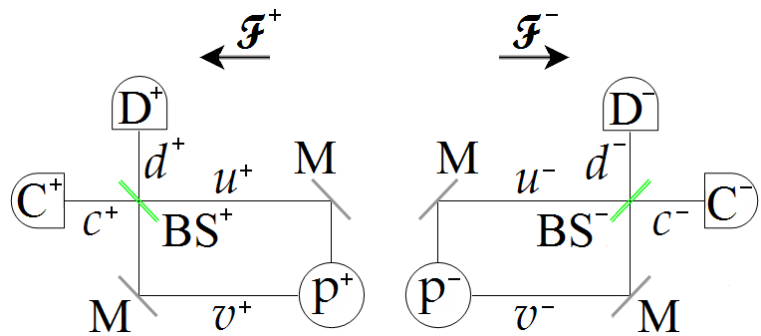

Figure 1. Hardy's thought experiment.

The quantum particles $p^+$ and $p^-$ have the possibility to follow the paths $u^+$, or $v^+$, respectively $u^-$, or $v^-$, conditioned by the correlations in the equation (1). M are mirrors. $BS^\pm$ are 50%-50% beam-splitters, and $C^\pm$, $D^\pm$ are ideal, absorbing detectors. $\mathcal{F}^+$ and $\mathcal{F}^-$ are frames of coordinates flying in the indicated directions.

$$|\psi\rangle_+ = \frac{i}{\sqrt{6}}\left[\,|c^+\rangle(i|u^-\rangle + 2|v^-\rangle) + |d^+\rangle|u^-\rangle\,\right]. \tag{3}$$

This state predicts that if $p^+$ triggers the detector $D^+$, $p^-$ could travel only on the path $u^-$.

By symmetry, according to the time axis of a frame of coordinates $\mathcal{F}^-$ flying in opposite direction than $\mathcal{F}^+$, by the time when $p^-$ is detected, $p^+$ didn't yet meet the beam-splitter $BS^+$. The calculi in [18] show that the state reflecting this situation is

$$|\psi\rangle_- = \frac{i}{\sqrt{6}}\left[\,(i|u^+\rangle + 2|v^+\rangle)|c^-\rangle + |u^+\rangle|d^-\rangle\,\right]. \tag{4}$$

This state predicts that if the detector $D^-$ is triggered, $p^+$ could travel only on the path $u^+$.

However, by the w-f $|\psi\rangle_0$, written according to the lab-frame, there is no state $|u^+\rangle|u^-\rangle$. There is either $|u^+\rangle|v^-\rangle$, or $|v^+\rangle|u^-\rangle$.

Recalling that one of the principles of the dBB mechanics is that particles follow continuous trajectories, one cannot escape the problem by suggesting that the two particles jump between the paths $u^+$ and $v^+$, respectively $u^-$ and $v^-$.

Some physicists are not convinced that the lack of relativistic covariance of the dBB trajectories invalidates the dBB interpretation. The common objection is that when one uses the relativity in combination with the quantum theory, one should work with the quantum field theory. However, as the velocities of the moving frames mentioned in [17] are low, there is no need to use the quantum field theory.

Though, trials were done to test the dBB mechanics reasoning within a single frame.

Englert, Scully, Süssman, and Walther described a thought-experiment, [19], the analysis of which placed a question mark on the expression of the dBB velocity. Starting from this expression they tried to derive a contradiction between the predictions of the dBB and QM mechanics. Their work triggered a debate to which participated both experimenters and theoreticians [20 – 24], however, all the participants whose articles were accessible to the present author, based their arguments on altered versions of the experiment, fact that rendered their conclusions arguable or non-relevant.

P. Ghose also proposed an experiment [25], implemented by Brida et al. [26], meant to show that the dBB velocity leads to predictions contradicting QM. An experiment akin to [25] was proposed by Golshani and Akhavan [27]. These proposals also gave birth to a debate [28 – 32]. The main criticism on Ghose's proposal was that the initial conditions under which the contradiction appears, occur in a minority of cases, and from the detection results one cannot infer the initial conditions. The debate seems not to have been concluded.

However, there is a point to be stressed vis-à-vis these debates. In the analysis in [18], the incompatibility between the dBB interpretation and the standard QM didn't come from one or another dBB equation. It came from a more general feature, the continuity of the dBB trajectories. All the class of theories which assume continuous trajectories for some substructure object should also disagree with the QM. To this class belongs also the hypothesis of 'empty/full waves' if the full waves are assumed to have continuous trajectories. This hypothesis is general, it does not propose particular formulas.

As the experiment described in [18] is a relativistic one, some physicists hold that the dBB mechanics can be considered correct in the non-relativistic domain. Vis-à-vis this claim, the section 3 of [33] presents a contradiction between the dBB mechanics and QM, reasoning within one single frame of coordinates. Its main line is described below.

The w-f exiting a source S is split by beam-splitters into three identical w-ps, $/a\rangle$, $/b\rangle$, and $/c\rangle$ – figure 2. On the path of each w-p is placed a phase-shifter. On each one of the end-beam-splitters, $BS_1$, $BS_2$, $BS_3$, land two inputs: one is a w-p from a source S, the other is an oscillator wave. The outputs impinge on two detectors.

The relevant trials of the experiment are those in which all the three detectors $D_1$, $D_2$, and $D_3$, click. The calculi in [33] showed that for the phase-shifts $\theta_1 = \theta_3 = 0$, $\theta_2 = \pi$, this result may be obtained in two ways:

1) the w-p $/a\rangle$ triggers the detector $D_1$, and the detectors $D_2$ and $D_3$ are triggered by the oscillators;

2) the w-p $/c\rangle$ triggers the detector $D_3$, and the detectors $D_2$ and $D_1$ are triggered by the oscillators.

Introducing now the concept of dBB particle guided by the w-f, i.e. traveling inside one of the w-ps, a self-contradictory situation appears:

i)   The total probability obtained for the joint detection in $D_1$, $D_2$, and $D_3$, is $\mathcal{M}^2/8$, where $\mathcal{M}$ is a normalization constant.

ii)  The probability that this joint detection be obtained in the case (1) is also $\mathcal{M}^2/8$. Therefore, for this type of detection to occur, the dBB particle from the source S should have traveled inside the w-p $/a\rangle$.

iii) The probability that this joint detection be obtained with the case (2) is also $\mathcal{M}^2/8$. Therefore, for this type of detection to occur, the dBB particle from the source S should have traveled inside the w-p $/c\rangle$.

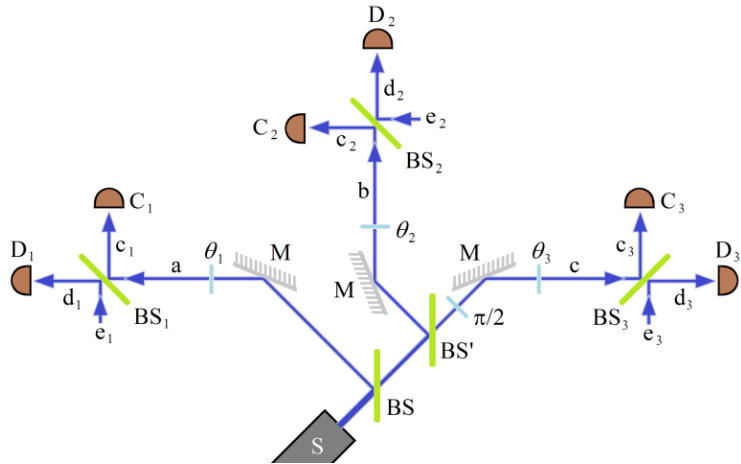

Figure 2. An arrangement for simultaneous detections of one particle in different places.

The colors are only for eye-guiding. A source S emits a one-particle w-f at low rate, so that at each time only one particle from S is present in the apparatus. The w-f is split by the beam-splitters BS and BS' into three identical w-ps $/a\rangle$, $/b\rangle$, and $/c\rangle$. The w-ps are re-oriented by mirrors to the end-beam-splitters $BS_1$, $BS_2$, $BS_3$, respectively. The w-p flying to $BS_n$ passes through the phase-shifter $\theta_n$. On $BS_n$ lands an additional input, $/e_n\rangle$, from a local oscillator. The outputs from $BS_n$ go to the detectors $D_n$ and $C_n$.

But that is impossible, one of the de Broglie-Bohm mechanics axioms is that the dBB particles follow continuous trajectories, do not jump between the w-ps $/a\rangle$ and $/c\rangle$.

## 2.2. The random discontinuous motion of particles

Following the objection of Born to Schrödinger in what concerns a charge distribution of an electron, at the 1927 Solvay conference [34], S. Gao realized [6] that

"there is no electrostatic self-interaction of the charge distribution of a quantum system"

Therefore, two wave-packets of the wave-function of an electron won't repel one another. After an examination of different solution to this situation he came to the opinion that at a given time, the charge is located in only one of the two wave-packets. He emitted the following view:

1) A single-component quantum system (e.g. an electron, a proton, an atom for which only the center-of-mass motion is studied, etc.) is localized at a given time in only one place.
Obviously, this view attributes to the quantum system a *particle* behavior;

2) The quantum system has a random, discontinuous movement (RDM), i.e. jumps from one position to another extremely frequently, the whole tableau of positions depicting the distribution predicted by the w-f:

"the ergodic motion of a particle is discontinuous, and the probability density that the particle appears in each position is equal to the modulus squared of its wave function there."

This interpretation is especially worth examining because it provides a simple justification to the following facts: a) the stochasticity of the results in measurements of quantum systems; b) from all the w-ps of a wave-function, only one triggers a detector; c) the w-ps of the w-f of a charged particle do not repel one another; d) in the Schrödinger equation of two particles, 1 and 2, we write the interaction potential in the form $V(\mathbf{r}_1 - \mathbf{r}_2, t)$, i.e. as if at the time $t$ the particle 1 is at the position $\mathbf{r}_1$ and the particle 2 is at the position $\mathbf{r}_2$.

On the other hand, Gao's interpretation has weaknesses. If the only ontology of this interpretation are particles moving discontinuously at random, it is not clear how can be prepared different w-fs; all the w-fs should be the same.

A serious problem deriving from Gao's interpretation is with entanglements. For instance, if two particles are entangled by positions, and each particle jumps at random times from one position to another, the correlations are bound to be violated. Gao proposed the solution of simultaneous jumps:

"Physical entity 1 . . . jumps discontinuously between positions $(x_1; y_1; z_1)$ and $(x_3; y_3; z_3)$, and physical entity 2 . . . jumps discontinuously between positions $(x_2; y_2; z_2)$ and $(x_4; y_4; z_4)$. Moreover, they jump in a precisely simultaneous way. When physical entity 1 jumps from position $(x_1; y_1; z_1)$ to position $(x_3; y_3; z_3)$, physical entity 2 always jumps from position $(x_2; y_2; z_2)$ to position $(x_4; y_4; z_4)$, and vice versa."

However, Gao didn't say which ontological entity *imposes* to the particles to jump in the way he postulated. Also, because of the relativity of simultaneity, in one frame of coordinates the jumps may be simultaneous, as Gao said, but in another frame there would be a delay between the jumps and a simultaneous measurement would find the correlations of the entanglement violated.
Gao was aware of this problem and admitted the existence of a preferred frame:

"In any case, whether simultaneity is relative or absolute, there is always a preferred Lorentz frame for the RDM of particles and its collapse evolution in the relativistic domain."

Regrettably, neither a preferred frame is a solution, because the experimenters may choose to test their particles at arbitrary, different times, which may turn out to be non-simultaneous in the preferred frame.

A problem appears even with single particle w-fs. For instance, if the w-f has two w-ps, A and B, and the particle jumps from A to B with non-superluminal velocity, for a while it won't be present in anyone of the w-ps. Neither could it be present in the intermediate space, where the w-f is null. If it jumps at superluminal velocity, a frame of coordinates can be found by which the particle jumps backwards in time, i.e. arrives at B before leaving A. Thus, for a while, the particle would be present in both w-ps.

## 2.3. The consistent histories

This interpretation is mainly due to R. Griffith. He proposed a modification of the quantum formalism by introducing so-called 'histories' of the evolution of quantum systems. The histories are not unitary evolutions, in contradiction with the quantum formalism by which a quantum system evolves unitarily before the macroscopic measurement. While claiming that the collapse hypothesis is no more necessary in this interpretation, Griffith introduced the collapse in the histories, as will appear in the example below. It will be proved that this procedure entails contradiction with quite trivial quantum predictions.

The figure 3 shows the configuration of one of the examples in [7]. On the beam-splitter, $BS_1$, of a Mach-Zehnder interferometer, lands the w-p of a photon (or of another microscopic particle). The w-p is split into two identical copies. They evolve, passing through stages indicated in the figure by circles. The inner arms of the interferometer are identical in length, and on them are inserted phase-shifters. The w-ps exit the interferometer through a second beam-splitter, $BS_2$. If the initial w-p lands on the side $a$ of $BS_1$, it evolves as follows:

$$/0a\rangle \to \frac{i/1c\rangle + /1d\rangle}{\sqrt{2}} \to \frac{i/2c\rangle + /2d\rangle}{\sqrt{2}} \to \frac{ie^{i\phi_c}/3c\rangle + e^{i\phi_d}/3d\rangle}{\sqrt{2}} \to ie^{i\alpha}\left[\sin(\beta)/4e\rangle + \cos(\beta)/4f\rangle\right],^2 \qquad (5)$$

where $\alpha = (\phi_d + \phi_c)/2$, and $\beta = (\phi_d - \phi_c)/2$.

For this evolution Griffiths proposed two histories, $Y$ and $Y'$, (relations (13.7 in chapter 13 of [7]). These histories consist in the same sequence of states as in (5) except for the final state. Except for a phase factor, $Y$ contains as final state only $/4e\rangle$, and $Y'$, only $/4f\rangle$. So, Griffiths assumed that the state of the quantum system before the measurement is not the coherent superposition in the rightmost expression in (5), but, the state $/e\rangle$ in some trials, and in the other trials, the state $/f\rangle$. He said that the probability of the history $Y$ ($Y'$) to occur, is equal to the absolute square of the amplitude of the state $/4e\rangle$ ($/4f\rangle$) in (5), i.e. $\sin^2(\beta)$ for final state $/e\rangle$, and $\cos^2(\beta)$ for final state $/f\rangle$.

However, a simple experiment can disprove these assumptions. Let, for instance $\beta = \pi/4$. Leaving aside leading phase factors, the final state in (5) would be a quantum superposition

---

[2] The numbers inside the states are the stages numbers, not number of particles. Mentioning the stage economizes writing the phase accumulated along the path to that stage, except for the phase-shifts $\phi_c$ and $\phi_d$.

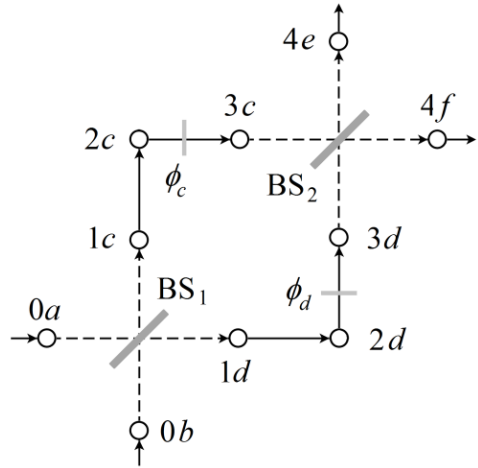

Figure 3. Consistent histories.

This figure is the main part of the figure 13.5 in the chapter 13 of [7]. The circles mark selected stages in the quantum system evolution. Their labels contain the stage number and the path. $BS_1$ and $BS_2$ are 50%-50% beam-splitters. $\phi_c$ and $\phi_d$ are phase-shifts.

$$|\psi\rangle_{\text{final}} \sim (|e\rangle + |f\rangle)/\sqrt{2}\,. \qquad (6)$$

Then, if the w-ps $|e\rangle$ and $|f\rangle$ are reflected by mirrors and cross one another before meeting detectors, in the crossing region will appear, after many trials, interference fringes. Such a tableau can't be produced without the coherent superposition in the last step in (5). If the last state in each trial is only $|e\rangle$, or only $|f\rangle$, in the crossing region would appear a featureless spot.

This interpretation is bound to disagree with the QM in additional issues. It assumes that a quantum system has at each time during its evolution, a certain definite property. In relation with such an assumption let's mention the famous dictum of A. Peres "*Unperformed experiments have no results*" [35].

On this ground A. Kent built objections [36] against the consistent histories interpretation showing that at a stage previous to the final one, a system may be with equal probability in one of two mutually orthogonal states, $|b\rangle$ and $|b'\rangle$. He called this situation 'contrary retrodiction'.

A controversy began between Griffith and Hartle on one side and Kent on the other side [37, 38]. I will not delve into it because the examples brought by Kent are no retrodiction at all. A retrodicted state $|x\rangle$ is a state preceding the final state of the system, and which is necessary for a certain result of the experiment to occur. The states $|b\rangle$ and $|b'\rangle$ of which spoke Kent were not necessary.

An example of retrodiction which indeed leads to an impossibility, is provided by the experiment in the figure 1 – see the explanation in the subsection 2.1. There are two retrodicted states $|u^+\rangle$ for the particle $p^+$, and $|u^-\rangle$ for the particle $p^-$, however, the wave-function contains no joint state $|u^+\rangle|u^-\rangle$.

Another example is provided by the experiment in the figure 2 – see the corresponding explanation. The trials in which the detectors $D_1$, $D_2$, and $D_3$, click together, impose contrary retrodictions for the state of the quantum system exiting the source S: 1) this quantum system was in the state $|a\rangle$; 2) this quantum system was in the state $|b\rangle$. Each one of these retrodictions implies a probability of the above triple detection, equal with its *total probability*. Therefore, in each trial in which the triple detection occurs, both states $|a\rangle$ and $|b\rangle$ occurred. But these states are orthogonal to one another, a quantum system cannot be in both of them at once.

Thus, the consistent histories interpretation makes inferences which are not consistent with the QM.

## 2.4. The transactions interpretation

Inspired by the Wheeler-Feynman absorber theory, J. G. Crammer proposed the hypothesis that both the emitter of a quantum system and the detector, emit a forward-in-time wave and a backward-in-time wave [8]. The detection is supposed to occur if 'hand-shake' occurs between the forward-in-time wave of the emitter and the backward-in-time wave of the detector. However, to the difference from the Wheeler–Feynman absorber theory in which the two waves superpose, in the formalism of the transaction interpretation appears the arithmetical product of the two waves. This replacement is motivated by claiming that the product gives the Born rule. It is a misunderstanding of the Born rule, the latter is the *inner* product of two waves (consisting in integration of the arithmetic product, over all its variables).

Besides that, it is not clear why for one detector the 'hand-shake' should occur, and for another detector shouldn't. For instance, if a w-p is split into two copies by a beam-splitter and each copy travels toward a separate detector, it is not clear what stops, in a given trial of the experiment, the hand-shake to happen at both detectors.

## 2.5. The 'many worlds' interpretation

In 1957, two years after Einstein's death, at Princeton university where had worked the father of the relativity theory, Hugh Everett exposed his dissertation on 'many universes' [9]. The main ideas are as follows: a macroscopic observer of a quantum system was supposed to have a microscopic state. When testing a single-particle w-f which is a quantum superposition of eigenstates of some operator $\hat{A}$, the observer multiplies into several ones. Each copy of the observer lives in another universe, and has a different microscopic state which gets correlated with one of the w-ps. Thus, each one of the w-ps triggers a detector, though in another world.

There are different oddities in this theory. First of all, the thesis contains no proof that Einstein's field equations imply a split of our universe into a multitude, at each measurement of a w-f which is a quantum superposition. Also, if the w-f is of the form

$$|\psi\rangle = (|a\rangle + \sqrt{2}\,|b\rangle)\,/\,\sqrt{3}\,, \tag{7}$$

the w-p $|b\rangle$ should impress a detector in our world twice as many times as the w-p $|a\rangle$, while in the other world the w-p $|a\rangle$ should impress a detector twice as many times than the w-p $|b\rangle$? Why this injustice?

Anyway, since we have no evidence of the existence of multiple universes, this theory remains fictive. Though, vis-à-vis the remark of S. Gao, from which one can infer that there can be no electrostatic interaction between at $|a\rangle$ and $|b\rangle$, it is difficult to banish the thought that if (in our world) the w-p $|a\rangle$ ($|b\rangle$) impresses a detector, the w-p $|b\rangle$ ($|a\rangle$) shows no effect, as if it does not exist.

## 3. The CSL model of collapse

It is a trivial fact that a microscopic system cannot be observed without unless it is amplified. However, with a quantum system the situation is more complicated, whatever we know about the system is a w-f. Many times this is a superposition of eigenstates of some operator, each eigenstate being characterized by an eigenvalue. We don't know what exactly impinges on the detectors, a wave or a particle. However, *only one* of the eigenvalues is reported, and if the respective detector is not an absorbing one, the experiment shows that from all the superposition only the corresponding eigenstate remain – what we cal 'collapse' of the w-f.

The CSL (continuous spontaneous localization) model of collapse assumes that the quantum formalism is correct as long as the system under study contains only a small number of components, however when the number of components increases, the Schrödinger equation has to be modified. When a quantum system impinges on a detector the impinging quantum system perturbs the components of the sensitive material in the detector – e.g. ionizes molecules – and the number of perturbed components increases due to secondary interactions up to macroscopic proportions, making the phenomenon observable.

The modification in the Schrödinger equation according to the CSL consists in two elements: 1) a noisy, stochastic term, which picks the eigenstate on which the w-f will collapse; 2) renormalization of the solution of the equation, rendering the equation nonlinear, and reducing the initial w-f to the picked eigenstate.

The fact that this model can simulate the collapse was exemplified in [15], however, not with complete success. The noise in the stochastic Schrödinger equation (SSE) is considered as varying quickly in time during the collapse. Not any sequence of noise values leads to a well-defined eigenstate from the initial superposition. Some sequences lead to non-physical solutions. Also, the probability of an eigenstate to be selected, has to be equal to the absolute square of the amplitude of that eigenstate in the initial w-f. Therefore the noise sequences have to be controlled by the initial w-f. Though, during the reduction process according to the SSE, the initial w-f is forgotten.

P. Pearle tried to justify by estimative considerations, [39], that the probabilities of the different eigenstates are respected by the CSL model. However, a big problem, escaping Pearle's shallow calculi, is that the solution of the SSE is highly dependent on the noise sequence and extremely unstable: a small modification in this sequence and the SSE solution may become non-physical. The problem was illustrated in [15].

Another question is the nature of this noise. It is clear that a classical field, even if noisy, won't fit the noise to each particular w-f, all the more, won't fit the noise to the nonlocal correlations in each entanglement.

The investigations with the CSL model are for the moment only in the beginning, mostly with single-particle w-fs. Thorough tests of the model on entanglements and in relativistic situations, are desirable. The present author intends to present in a future work *very serious problems* of this model vis-à-vis the relativity, of which the supporters of the model seem to be unaware at present.

D. Bedingham, [14], tried to apply the CSL model to the polarization singlet and predict the well-known probabilities for the various results, however, the procedure he used is alien to the CSL model. The collapse to the different possible states and with the corresponding probabilities should be due to the sequence of values of the stochastic noise during the measurement. Bedingham, however, introduced additional variables for obtaining the desired probabilities, variables correlated nonlocally in time and space, which do not belong to the CSL model. Also, his calculi predicted the effect of collapse for a very long time of measurement, instead of a high number of participating particles.

## 4. The collapse principle is unavoidable

The purpose of the present section is to prove that despite the fact that the collapse process cannot be described with the QM formalism, the collapse principle is unavoidable in the QM. Two theorems will be proved, illustrating the fact that denying the collapse one comes to predictions contradicting the QM. In proving both of them, the detectors will be considered as absorbing and ideal.

How can a detector remain non-impressed by a w-p?
Many answers can be proposed, and it is impossible to deal with all the suggestions the imagination can advance, moreover, some suggestions may contradict laws of physics, therefore they will be ignored. Two answers seem plausible to the present author:

1) The perturbation produced by the w-p in the detector is too small for changing the macroscopic state of the material in the detector.

2) The w-p 'does not feel' the detector, and passes by it without producing perturbation.

The following theorems rule out these options.

***Theorem 1:***
*If a w-p meets a detector without triggering it, no particle in the detector remains perturbed.*

***Proof:***
From a down-conversion pair of photons, the idler photon is sent to a detector Q – figure 4. The w-p of the signal photon is split by the 50%-50% beam-splitter BS into two copies, one reflected, $|a\rangle$, and one transmitted, $|b\rangle$. Both copies travel toward a rotating mirror M, initially in horizontal position, so that when $|a\rangle$ touches it, the w-p is reflected toward the detector S. The path from the nonlinear crystal (not shown in the figure) which produces the pair, to the detector Q, and the path to the mirror M, are tuned so that the idler reaches and triggers Q, immediately after $|a\rangle$ was reflected by M. The w-p $|b\rangle$ reaches the region of the mirror M later than $|a\rangle$ because of a retarding system of mirrors m. Upon the click of Q, M is rotated to vertical position, therefore $|b\rangle$ does not meet this mirror and continues its travel toward the detector S on the same track as $|a\rangle$. At this step the two w-ps look as in the figure, and the w-f of the signal photon becomes, considering all the reflections at mirrors and the additional phase due to the path-length difference,

$$|\phi\rangle = -(\,e^{i\varphi}|a\rangle + i|b\rangle\,)\,/\sqrt{2}\,. \tag{8}$$

When the w-p $|a\rangle$ enters the detector, it interacts with a first few particles (atoms/molecules) from the sensitive material in the detector, so that an entanglement is generated

$$|\phi\rangle|A_1^{u}\rangle|A_2^{u}\rangle\ldots|A_n^{u}\rangle \rightarrow -(\,e^{i\varphi}|a^{p}\rangle|A_1^{p}\rangle|A_2^{p}\rangle\ldots|A_n^{p}\rangle + i|b^{u}\rangle|A_1^{u}\rangle|A_2^{u}\rangle\ldots|A_n^{u}\rangle\,)\,/\sqrt{2}\,, \tag{9}$$

where the upper-script 'p' stands for 'perturbed', and 'u' for 'unperturbed'.
Further, the perturbed particles meet and perturb additional particles. As long as the number of perturbed particles is small, the total system continues to be described by the quantum formalism,

$$|\Phi\rangle = -(\,e^{i\varphi}|a^{p}\rangle|A_1^{p}\rangle|A_2^{p}\rangle\ldots|A_N^{p}\rangle + i|b^{u}\rangle|A_1^{u}\rangle|A_2^{u}\rangle\ldots|A_N^{u}\rangle\,)\,/\sqrt{2}\,. \tag{10}$$

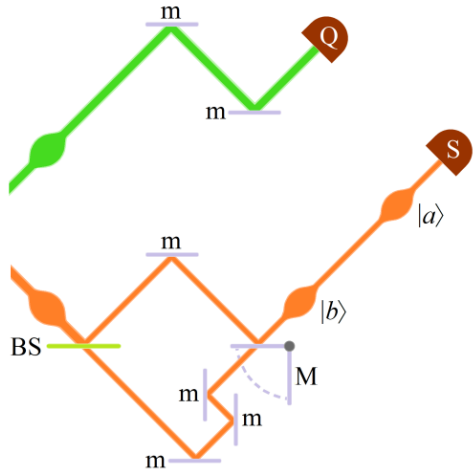

Figure 4. A which-way experiment.

The colors in the figure are only for eye-guiding. BS is a beam-splitter, m are fixed mirrors, M is a rotatable mirror, S and Q are ideal detectors. From a down-conversion pair, the idler photon (green) travels toward the detector Q, and the signal photon (orange) lands on a fair beam-splitter BS, where its w-p is split into a reflected part $|a\rangle$, and a transmitted part $|b\rangle$. The mirror M is initially in horizontal position, thus, $|a\rangle$ is reflected by M and directed toward the detector S. Immediately after that, the idler photon meets the detector Q which clicks. Upon this event, M is rotated to vertical. So, the w-p $|b\rangle$ retarded by three mirrors m, doesn't encounter the mirror M and travels towards the detector S, along the same track as $|a\rangle$.

At some time $t$ the w-p $|b\rangle$ also reaches the detector S. Let the state of the total system at this time be as described by the equation (10). $|b\rangle$ also interacts with the sensitive material in the detector perturbing in the beginning, say, $k$ particles, which in continuation would perturb other particles.

Assume for simplicity that the two sets of perturbed particles are disjoint. If so, the equation (10) evolves into

$$|\Theta\rangle = -\left(e^{i\varphi}/a^p\rangle|A_1^p\rangle\ldots|A_N^p\rangle|B_1^u\rangle\ldots|B_K^u\rangle + i|b^p\rangle|A_1^u\rangle\ldots|A_N^u\rangle|B_1^p\rangle\ldots|B_K^p\rangle\right)/\sqrt{2}\,. \tag{11}$$

Let's see what would be the meaning of the RHS of (11) if the numbers $N$ and $K$ would increase to *macroscopic* values.

One can see in the second term between the round parentheses that if the set of particles $\{B\}$ are perturbed, then the set $\{A\}$ should *not* remain perturbed. That means, no macroscopic perturbation – avalanche – can unfold due to the w-p $|a\rangle$. In fact, one cannot have even $|A_1^p\rangle$, i.e. not even one particle from the set correlated with $|a\rangle$ can remain perturbed.

Symmetrically, if the set$\{A\}$ are perturbed, then the set $\{B\}$ is forced to remain unperturbed, therefore no avalanche can begin from them. In particular, one cannot have even $|B_1^p\rangle$, i.e. not even one particle from the set correlated with $|b\rangle$ can remain perturbed.

That confirms the theorem and rules out the option (1).

For clarity, let's say again that if the sets $\{A\}$ and $\{B\}$ are small – a few particles – the entanglement is valid. Only if one set increases to macroscopic population, then the other set leaves no effect, $K = 0$ ( $N = 0$).

In fact, behind the proof of this theorem stands a trivial quantum fact: the initial w-f is a one-particle w-f, it contains no product $/a\rangle/b\rangle$, therefore it cannot leave in the detector S perturbations from both w-ps.

***Theorem 2:***
*A w-p that does not trigger an ideal, absorbing detector, is destroyed.*

***Proof:***
The following experiment will be performed in two steps.

**A)** A source S of single-particle beams emits w-ps which pass through the beam-splitters BS and BS' – figure 5. Three identical w-ps, $/\mathbf{a}\rangle$, $/\mathbf{b}\rangle$, $/\mathbf{c}\rangle$, are generated. All the w-ps are reflected so as to cross one another simultaneously in the same region I. On the paths of $/\mathbf{a}\rangle$ and $/\mathbf{c}\rangle$ are inserted phase-shifts. In all, the w-f is

$$/\psi\rangle = -\frac{1}{\sqrt{3}}\left(e^{i\theta_1}/\mathbf{a}\rangle + /\mathbf{b}\rangle + e^{i\theta_3}/\mathbf{c}\rangle\right).\tag{12}$$

In the region I is inserted a sensitive plate perpendicularly to the direction of flight of $/\mathbf{b}\rangle$. Thus, after many trials, an interference pattern is formed on it. The position of the fringes depends on the phase-shifts $\theta_1$ and $\theta_3$. The beam $/\mathbf{b}\rangle$ falls on the plate in the direction parallel with the fringes, introducing a background on which are superimposed the fringes.

The interference pattern proves that in absence of any perturbation of the w-ps, all the three w-ps are present in the apparatus.

**B)** On the path of the w-p $/\mathbf{b}\rangle$, before entering the region I, we place a detector, E – not shown in the figure. Obviously, in 1/3 of the trials this detector clicks, and in the rest, it doesn't.

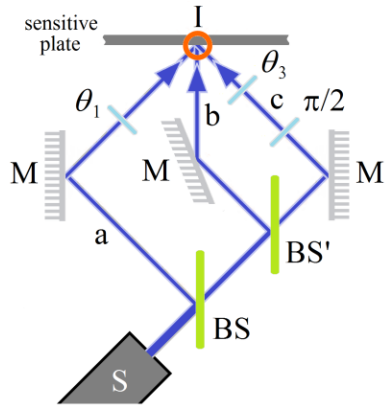

Figure 5. Interference of three wave-packets.
The colors are only for eye-guiding. S is a source which emits particles at slow rate so as at any time at most one particle travels in the apparatus. BS and BS' are beam-splitters, BS reflects 1/3 from the incident beam intensity, and BS' reflects and transmits in equal proportion. Thus, they separate the initial w-p into three identical copies. M are mirrors, $\theta_1$ and $\theta_3$ are phase-shifts. The additional phase-shift of $\pi/2$ compensates the phase of the w-p $/\mathbf{c}\rangle$ which is reflected only once, while $/\mathbf{a}\rangle$ and $/\mathbf{b}\rangle$ have been reflected twice. The red circle surrounds the triple interference region, I.

The QM predicts that in the trials in which E clicks, the sensitive plate is not impressed. It is clear that the absorbing detector E absorbed the w-p $/\mathbf{b}\rangle$, but what about $/\mathbf{a}\rangle$ and $/\mathbf{c}\rangle$? In the experiment **(A)** we saw that in the absence of the detector E, $/\mathbf{a}\rangle$ and $/\mathbf{c}\rangle$ were present.

So, it seems that when $/\mathbf{b}\rangle$ is detected, the w-ps $/\mathbf{a}\rangle$ and $/\mathbf{c}\rangle$ disappear, or, are destroyed by the sensitive plate without leaving on it any signature, as says the theorem 1.

In the trials in which E doesn't click, the QM predicts that the plate is impressed. After many trials an interference pattern appears as in the experiment **(A)**, however, *without the background*, i.e. without the effect of $/\mathbf{b}\rangle$. Let's remind again that the experiment **(A)** proved that without the presence of the detector, the w-p $/\mathbf{b}\rangle$ reached the plate.

Therefore, it seems that although $/\mathbf{b}\rangle$ impinged on the detector, if the detector is not impressed $/\mathbf{b}\rangle$ disappeared, or is destroyed by the detector as says the theorem 1.

If, as supposed the option (2) in the beginning of this section, $/\mathbf{b}\rangle$ just ignores the detector in some way (not clear to us) and reaches the plate, it should perturb the interference pattern adding a background, as results from the w-f (12). So, the option (2) is ruled out.

Bringing together all the above inferences, the answer to the question asked in the beginning of this section is that the detection of one w-p entails the disappearance or destruction of the other wave-packets, and the non-detection of a w-p leaves unperturbed the other w-ps. This is what we understand by 'collapse' of the w-f.

## 4. Collapse without a macroscopic measurement is a misunderstanding

The Einstein-Podolsky-Rosen (EPR) thought-experiment [40] proved that the QM does not allow what they called 'elements of reality'

> "If without in any way disturbing a system, we can predict with certainty the value of a physical quantity, then there exists an element of physical reality corresponding to this physical quantity"

The examination of the EPR thought-experiment is directly connected to the issue of the collapse, though, in a way not discussed in the previous sections.

Let's remind the essence of this thought-experiment: two quantum particles, I and II, are entangled by both position and linear momentum, in a way that by measuring the linear momentum of the particle I one can predict with certainty the linear momentum II, and measuring the position of II one can predict with certainty the position of I.

However, for each particle separately, the operators of position and linear momentum don't commute, so that the system can't have both of them well defined.

EPR proposed two solutions for this incompatibility:

> "From this follows that either (1) *the quantum-mechanical description of reality given by the wave function is not complete* or (2) *when the operators corresponding to two physical quantities do not commute the two quantities cannot have simultaneous reality.*"

From the experience accumulated along tens of years after the EPR article, it appears that the QM predictions were never proved wrong. Attempts to modify the QM formalism, in the section 2 was proved for the most popular interpretations that they failed. The QM formalism seems to be immune against modifications. Thus, until now, the proposal (1) proved to be useless.

As to the proposal (2), A. Peres gave a clearer formulation, [35]: "*unperformed experiments have no results*". That means, the linear momentum found by testing the particle I predicts nothing for the particle II, unless the latter also undergoes a test of linear momentum, and analogously, the test of the particle II position predicts nothing for the particle I.

The experiment in [17] and the analysis in [18] lead to the same conclusion. Let's remind briefly the relevant steps – see section 2.1, figure 1. The quantum particles $p^+$ and $p^-$, entangled by paths of flight, pass through beam-splitters beyond which the particles are detected. The relevant case is the joint detection in the detectors $D^+$ and $D^-$. If the 'collapse at a distance' were valid, judging according to the moving frame of coordinates $\mathcal{F}^+$, the detection by $D^+$ should entail the collapse of the particle $p^-$ on the state $/u^-\rangle$. Symmetrically, according to the moving frame of coordinates $\mathcal{F}^-$, the detection by $D^-$ should entail the collapse of $p^+$ on the state $/u^+\rangle$. However, the w-f contains no component $/u^+\rangle/u^-\rangle$.

## 6. Conclusions

This work proved that the most popular interpretations of the QM which deny the principle of collapse of the w-f, make predictions which in some experiments disagree with the predictions of the QM. Since never an experiment performed on quantum systems produced results that contradicted the QM predictions, it can be concluded that these interpretations failed.

Despite of that, the present author wishes to call the attention on the proposal of S. Gao [6] – subsection 2.2. As the experiments prove, different parts of the w-f of a charged particle do not repel one another; though, an electric or magnetic field deflects the parts as if each one of them carries the entire electric charge. For instance, if the w-f consists in $N$ w-ps, they are deflected as if the charge was multiplied into $N$ copies, one per w-p. S. Gao proposed the idea of a charged particle jumping at random between the different parts of the w-f. As shown in the subsection 2.2, this idea is problematic vis-à-vis entanglements and relativity. Though, the present author considers it as worth of deep examination and, hopefully, improvements.

Two proposals which take the collapse as a real phenomenon are the GRW and the CSL models of collapse. In this work only the CSL model was discussed from the two, as it seems to the present author, better formulated mathematically. As shown in [15] the model is able to explain, step by step, the collapse process unfolding in a detector. However, as said in section 3, there are problems with the model. The SSE predicts sometimes non-physical solutions, and although the model reached the age of 30 years, it is not yet clear whether it can explain entanglements and whether it is compatible with the relativity. The present author works on a material which points to very severe problems in these two situations.

As the collapse process stands at the crossroads between the quantum and the macroscopic world, it cannot be treated with the QM formalism. However, it passed unnoticed until now that this formalism *needs* the collapse principle. Two theorems were proved here, showing that in the absence of the collapse, effects contradicting the QM should appear.

On the other hand, a misconception widely spread in the quantum community was attacked: it is believed by many physicists that in entangled systems, the measurement of one of the systems with a given result, collapses the other system(s) participating in the entanglement, to a well-defined state – see for instance [41]. In the famous EPR article [40] it was shown that such a belief leads to a contradiction with the rules of the QM. An article of Berndl and Goldstein [42] which explained the root of the so-called Hardy's paradox [17], also indicated quite clearly that it is impossible to infer from the measurement of one particle, in which state is left another particle, entangled with the former. It's worthy to cite again Peres' dictum "*unperformed measurements have no results*". The QM predicts only results of measurements which are going to be actually performed.

---

# References

[1]  J. von Neumann, "*Mathematische Grundlagen der Quantenmechanik*", Springer, Berlin, (1932); English translation by R. T. Beyer, "*Mathematical Foundations of Quantum Mechanics*", Princeton University Press, Princeton, N. J., (1955).

[2]  G. Lüders, "*Über die Zustandsänderung durch den Meßprozeß*", Annalen der Physik **443**, issue 5-8, pages 322-328 (1950); English translation and discussion by K. A. Kirkpatrick, "*Concerning the state-change due to the measurement process*", Ann. Phys. (Leipzig) **15**, issue 9, pages 663-670 (2006), arXiv:quant-ph/0403007v2.

[3]  L. de Broglie, "*Ondes et mouvements*", publisher Gauthier-Villars, (1926); "*An introduction to the study of the wave mechanics*", translation from French by H. T. Flint, D.Sc, Ph.D., first edition 1930.

[4]  David Bohm, "*A suggested interpretation of the quantum theory in terms of "hidden" variables*", parts I and II, Phys. Rev. **85**, pages 166-179, respectively pages 180-193 (1952).

[5]  L. Hardy, "*On the existence of empty waves in quantum theory*", Physics Letters A, vol. **167**, issue 1, pages 11-16 (1992).

[6]  S. Gao, "*MEANING OF THE WAVE FUNCTION - In search of the ontology of quantum mechanics*", arXiv:quant-ph/arXiv:1611.02738v1

[7]  R. B. Griffiths, "*Consistent histories and the interpretation of quantum mechanics*", J. Stat. Phys. **36**, pages 219-272 (1984); idem, "*Consistent quantum theory*", Cambridge, U.K.: Cambridge University Press (2002); "*The Consistent Histories Approach to Quantum Mechanics*", Stanford Encyclopedia of Philosophy, (First published Thu Aug 7, 2014; substantive revision Thu Jun 6, 2019).

[8]  J. G. Cramer, "*The Transactional Interpretation of the Quantum Mechanics*", Rev. Mod. Phys. **58**, no. 3, pages 647-688, (1986); "*An Overview of the Transactional Interpretation*", Int. J. of Th. Phys. **27**, no. 2, pages 227-236 (1988).

[9]  H. Everett, "*The theory of the Universal Wavefunction*", Thesis, Princeton University, pages 1-140 (1956, 1973).

https://www.researchgate.net/publication/280676329_A_New_Ontological_Interpretation_of_the_Wave_Function; idem, "*Meaning of the wave function: In search of the ontology of quantum mechanics*", arXiv:quant-ph/1611.02738v1.

[10] R. P. Feynman and A. R. Hibbs, "*Quantum Mechanics and Path Integral*", McGraw-Hill Companies, Inc., New York, (1965); emended edition Daniel F. Styer (2005); emended re-publication Dover Publication, Inc., Mineola, New York (2010).

[11] G-C. Ghirardi, A. Rimini and T. Weber, "*Unified dynamics for microscopic and macroscopic systems*", Phys. Rev. D **34**, page 470 (1986).

[12] G-C Ghirardi, P. Pearle, and A. Rimini, "*Markov processes in Hilbert space and continuous spontaneous localization of systems of identical particles*", Phys. Rev. A: Atomic, Molecular, and Optical Physics **42**, no. 1, pages 78-89 (1990).

[13] A. Bassi and G-C. Ghirardi, "*Dynamical Reduction Models*", Phys. Rept. **379**, page 257 (2003).

[14] D. J. Bedingham, "*Dynamical state reduction in an EPR experiment*", J. Phys. A: Mathematical and Theoretical **42**, 465301 (2009).

[15] S. D. Wechsler, "*In praise and in Criticism of the Model of Continuous Spontaneous Localization of the Wave-Function*", to appear in JQIS **10**, no. 4, (2020).

[16] K. Berndl, S. Goldstein, and N. Zanghì, "*EPR-Bell Nonlocality, Lorentz Invariance, and Bohmian Quantum Theory*", Phys. Rev. A **53**, page 2062, (1 April 1996), arXiv:quant-ph/9510.027 .

[17] L. Hardy, "*Quantum Mechanics, Local Realistic Theories, and Lorenz-Invariant Realistic Theories*", Phys. Rev. Lett. **68**, no. 20, page 2981 (1992).

[18] S. Wechsler, "*Hardy's paradox made simple – what we infer from it?*" https://www.researchgate.net/publication/318446904_Hardy's_paradox_made_simple_-_what_we_infer_from_it (2017)

[19] B-J. Englert, M. O. Scully, G. Süssman, and H. Walther, "*Surrealistic Bohmian Trajectories*", Z. Naturforsch **47a**, pages 1175-1186 (1992).

[20] C. Dewdney, L. Hardy, E.J. Squires, "How late measurements of quantum trajectories can fool a detector", Physics Letters A **184**, 6, (1993).

[21] D. H. Mahler, L. Rozema, K. Fisher, L. Vermeyden, K. J. Resch, H. M. Wiseman, A. Steinberg, "*Experimental nonlocal and surreal Bohmian trajectories*", Sci. Adv. 2016; 2 : e1501466, http://advances.sciencemag.org/.

[22] Detlef Dürr, Walter Fusseder, Sheldon Goldstein, and Nino Zanghì, "*Comment on "Surrealistic Bohm Trajectories"* ", Z. Naturforsch. **48a**, 1261, (1993).

[23]  B-G. Englert, M. O. Scully, G. Süssmann, H. Walther, "*Reply to Comment on "Surrealistic Bohm Trajectories"* ", Z. Naturforsch. **48a**, 1263, (1993);

[24]  B. J. Hiley and R. E. Callaghan, "*Delayed Choice Experiments and the Bohm Approach*", arXiv:1602.06100v1.

[25] P. Ghose, "*On the Incompatibility of Standard Quantum Mechanics and the de Broglie-Bohm Theory*", arXiv:quant-ph/0103126.

[26] G. Brida et al., "*A biphotons double slit experiment*", Phys. Rev. A **68**, 033803 (2003).

[27] M. Golshani and O. Akhavan, "*Experiment can decide between standard and Bohmian quantum mechanics*", quant-ph/0103100; idem, J. Phys. A **34** pages 5259-5268 (2001), arXiv:quant-ph/0103101.

[28] L. Marchildon, "*On Bohmian trajectories in two-particle interference devices*", arXiv:quant-ph/0101132.

[29] P. Ghose, "*Comments on "On Bohm trajectories in two-particle interference devices" by L. Marchildon*", arXiv:quant-ph/0102131.

[30] W. Struyve and W. De Baere, "*Comments on some recently proposed experiments that should distinguish Bohmian mechanics from quantum mechanics*", arXiv:quant-ph/0108038v1.

[31] P. Ghose, "*Comments on Struyve and Baere's paper on experiments to distinguish Bohmian mechanics from quantum mechanics*", arXiv:quant-ph/0208192.

[32] M. Golshani and O. Akhavan, "*On the Experimental Incompatibility Between Standard and Bohmian Quantum Mechanics*", arXiv:quant-ph/0110123.

[33] S. D. Wechsler, "*The wave-particle duality – Does the Concept of Particle Make Sense in Quantum Mechanics? Should we ask the Second Quantization?*", J. of Quantum Inf. Sci. **9**, no. 3 (September) pages 155-170 (2019).

[34]  G. Bacciagaluppi and A. Valentini, "Quantum Theory at the Crossroads: Reconsidering the 1927 Solvay Conference", Cambridge: Cambridge University Press, page 426 (2009).

[35]  A. Peres, "*Unperformed experiments have no results*", Am. J. Phys., **46**(7), pages 745-747 (1978).

[36] A. Kent, "*Consistent Sets Yield Contrary Inferences in Quantum Theory*", Phys. Rev. Lett. **78**, pages 2874-2877 (1997), arXiv:gr-qc/9604012v2.

[37] R. B. Griffith and J. B. Hartle, "*Comment on "Consistent Sets Yield Contrary Inferences in Quantum Theory*", Phys. Rev. Lett. **81**, pages 1981-1982 (1998)

[38] A. Kent, "*Consistent Sets and Contrary Inferences: Reply to Griffiths and Hartle*", Phys. Rev. Lett. **81**, page 1982 (1998).

[39] P. Pearle, "*Combining stochastic dynamical state-vector reduction with spontaneous localization*", Phys. Rev. A **39**, pages 2277-2289 (1989).

[40] A. Einstein, B. Podolsky, and N. Rosen, "*Can Quantum Mechanical Description of Physical Reality Be Considered Complete?*", Phys. Rev. **47**, no. 10, pages 777-780 (1935).

[41] Y. Xiao, Y. Kedem, J-S. Xu, C-F. Li, G-C. Guo, "*Experimental nonlocal steering of Bohmian trajectories*", Optics Express **25**, no. 13, pages 14463-14472 (2017).

[42] K. Berndl and S. Goldstein, "*Comment on "Quantum Mechanics, Local Realistic Theories, and Lorenz-Invariant Realistic Theories*", Phys. Rev. Lett. **72**, no. 5, pages 780-780 (1994).